\documentclass{ws-ijmpa}

\usepackage{amssymb}

\begin{document}

\markboth{A.~Krassnigg, C.\,D.~Roberts, and S.\,V.~Wright}
{Meson spectroscopy and properties using Dyson-Schwinger equations}

\catchline{}{}{}{}{}
%
%%%%%%%%%%%%%%%%%%%%%%%%%%%%%%%%%%%%%%%%%%%%%%%%%%%%%%%%%%%%%%%%%%%%

\title{%
$\;$\,\\[-6ex]
\hspace*{\fill}{\tt\normalsize ANL-PHY-11578-TH-2006}\\[1ex] %
Meson spectroscopy and properties using Dyson-Schwinger equations}

\author{A. KRASSNIGG}

\address{Institut f\"ur Physik, University of Graz, Universit\"atsplatz 5\\
A-8010 Graz, Austria\\
andreas.krassnigg@uni-graz.at}

\author{C.\,D.~ROBERTS and S.\,V.~WRIGHT}

\address{Physics Division, Argonne National Laboratory, 97000 South Cass Avenue\\
Argonne, IL 60439, USA\\
cdroberts@anl.gov and svwright@anl.gov}

\maketitle

\begin{history}
\received{15 08 2006}
\revised{Day Month Year}
\end{history}

\begin{abstract}
We study pseudoscalar and scalar mesons using a practical and symmetry preserving truncation
of QCD's Dyson-Schwinger equations.
We investigate and compare properties of ground and radially excited meson states.
In addition to exact results for radial meson excitations we also present results for meson
masses and decay constants from the chiral limit up to the charm-quark mass, e.\,g., the mass
of the $\chi_{c0}(2P)$ meson.

\keywords{Dyson-Schwinger equations; Bethe-Salpeter equation; meson radial excitations.}
\end{abstract}

\ccode{PACS numbers:%
%12.39.Ki% Relativistic quark model
%,21.45.+v% Few-body systems
%12.38.Aw% General properties of QCD (dynamics, confinement, etc.)
%,11.30.Rd% Chiral symmetries
,12.38.Lg% Other nonperturbative calculations (i.e. other than lattice)
,12.40.Yx% Hadron mass models and calculations
,11.10.St% Bound and unstable states; Bethe-Salpeter equations
}

\section{Introduction}

Dyson-Schwinger equations (DSEs) are a nonperturbative continuum approach to quantum
chromodynamics (QCD).\cite{Roberts:1994dr} They provide a means to study properties of the
Green functions of QCD\cite{Alkofer:2000wg,Fischer:2006ub} as well as hadrons as bound
states of quarks and gluons.\cite{Roberts:2000aa,Maris:2003vk,Holl:2006ni} Hadrons are
studied in this framework of infinitely many coupled integral equations with the help
of a symmetry preserving truncation scheme. In the case of mesons discussed here, one solves the Bethe-Salpeter
equation (BSE) for a quark-antiquark pair. For calculations of baryon properties,
e.\,g.~electromagnetic, weak, and pionic form
factors\cite{Hellstern:1997pg,Bloch:2003vn,Alkofer:2004yf,Holl:2005zi,Holl:2006zw}
one uses a covariant set of Faddeev equations.\cite{Hellstern:1997pg,Oettel:1998bk,Oettel:1999gc}

The Dyson-Schwinger equation framework has numerous features, amongst them: first, it is a
Poincar\'{e}-covariant framework and thus ideally suited for the study of hadron observables
such as, e.\,g., electromagnetic form factors. Secondly, symmetries are represented by
Ward-Takahashi or Slavnov-Taylor identities, which are then built into the scheme used
to truncate the infinite tower of coupled integral equations. If a truncation scheme respects
such an identity at every step, then one can i) prove exact (model-independent) results and ii)
use sophisticated models to calculate physical quantities which illustrate these results and automatically
reflect the properties of the corresponding symmetry.

One such truncation is the so-called rainbow-ladder truncation, which has been used extensively
and successfully to study meson ground states for more than a decade (see, e.\,g.,
Refs.~\refcite{Maris:1997hd,Maris:1997tm,Maris:1999nt,Maris:1999bh,Maris:2000sk}).
Despite this success it has become obvious that certain states and phenomena, such as axial-vector
mesons, exotic mesons,\footnote{We use the term ``exotic'' to characterize mesons with quantum
numbers that a system composed of a constituent-quark and constituent antiquark cannot have.} or heavy-light systems
are not well-described in the most sophisticated calculations available to date.
As a consequence, efforts are being made to go beyond this
truncation,\cite{Bender:1996bb,Bhagwat:2003vw,Bhagwat:2004hn,Watson:2004kd,Fischer:2005wx}
but these efforts are considerable, forcing present sophisticated calculations to remain at
an exploratory stage.

\section{QCD Gap and Bethe-Salpeter Equations}
\label{gapbse}

The homogeneous BSE is\footnote{We employ a Euclidean metric, with:
$\{\gamma_\mu,\gamma_\nu\} =
2\delta_{\mu\nu}$; $\gamma_\mu^\dagger = \gamma_\mu$; $ \gamma_5 = - \gamma_1
\gamma_2 \gamma_3 \gamma_4$; ${\rm tr}\, \gamma_5 \gamma_\mu
\gamma_\nu \gamma_\rho \gamma_\sigma = -4\, \varepsilon_{\mu\nu\rho\sigma}$;
and $a \cdot b = \sum_{i=1}^4 a_i b_i$.  For a timelike vector
$P_\mu$, $P^2<0$.}
\begin{equation}
\label{bse}
[\Gamma(p;P)]_{tu} = \int^\Lambda_q [\chi(q;P)]_{sr}\, K_{rs}^{tu}(p,q;P)\,,
\end{equation}
where $p$ is the relative and $P$ the total momentum of the constituents,
$r$,\ldots,\,$u$ represent color, Dirac and flavor indices,
\begin{equation}
\label{definechi}
\chi(q;P)= S(q_+) \Gamma(q;P) S(q_-)\,,
\end{equation}
$q_\pm = q\pm P/2$, and $\int^\Lambda_q$ represents a Poincar\'e invariant
regularization of the integral, with $\Lambda$ the regularization mass-scale.
In Eq.\,(\ref{bse}), $S$ is the renormalized dressed-quark propagator and $K$ is
the fully amputated dressed-quark-antiquark scattering kernel; for details,
see Refs.~\refcite{Maris:1997hd,Maris:1997tm}.

The dressed-quark propagator appearing in the BSE's kernel is determined by
the renormalized gap equation
\begin{eqnarray}
S(p)^{-1} & =&  Z_2 \,(i\gamma\cdot p + m^{\rm bm}) + \Sigma(p)\,, \label{gap} \\
\Sigma(p) & = & Z_1 \int^\Lambda_q\! g^2 D_{\mu\nu}(p-q) \frac{\lambda^a}{2}\gamma_\mu S(q)
\Gamma^a_\nu(q,p) , \label{sigma}
\end{eqnarray}
where $D_{\mu\nu}$ is the dressed gluon propagator, $\Gamma_\nu(q,p)$ is the
dressed quark-gluon vertex, and $m^{\rm bm}$ is the $\Lambda$-dependent current-quark bare mass.
The quark-gluon-vertex and quark wave function renormalization constants, $Z_{1,2}(\zeta^2,\Lambda^2)$,
depend on the gauge parameter, the renormalization point, $\zeta$, and the regularization
mass-scale. The leptonic decay constant of a pseudoscalar meson is calculated from the
solution of Eq.~(\ref{bse}) via
\begin{eqnarray}
\label{fpi} f_{\mathrm{PS}} \,\delta^{ij} \,  P_\mu &=& Z_2\,{\rm tr} \int^\Lambda_q
\frac{1}{2} \tau^i \gamma_5\gamma_\mu\, \chi^j_{\mathrm{PS}}(q;P) \,,
\end{eqnarray}
which is gauge invariant, and cutoff and renormalisation-point independent.

\subsection{Rainbow-ladder truncation}

The first step in the symmetry-preserving truncation scheme described in
Refs.~\refcite{Bender:1996bb,Bhagwat:2004hn,Munczek:1994zz,Bender:2002as} is the
rainbow approximation to the gap equation combined with a ladder truncation in the BSE.
The interaction kernels of Eqs.~(\ref{bse}) and (\ref{gap}) then take the form
\begin{eqnarray}
K^{tu}_{rs}(p,q;P) =   - \,4\pi\alpha(Q^2) \, D_{\mu\nu}^{\rm free}(Q)\,
\left[\gamma_\mu \frac{\lambda^a}{2}\right]_{ts} \, \left[\gamma_\nu \frac{\lambda^a}{2}\right]_{ru} \label{ladderK}
\end{eqnarray}
and
\begin{equation}
\Sigma(p)=\int^\Lambda_q\! 4\pi\alpha(Q^2) D_{\mu\nu}^{\rm free}(Q)
\frac{\lambda^a}{2}\gamma_\mu S(q) \frac{\lambda^a}{2}\gamma_\nu \;,  \label{rainbowdse}
\end{equation}
where $Q=p-q$, $D_{\mu\nu}^{\rm free}(Q)$ is the free gauge boson
propagator\footnote{We use Landau gauge in all calculations.} and
$\alpha(Q^2)$ is an effective running coupling. The ultraviolet behavior of this
coupling can be taken from perturbative QCD, while in the infrared one makes an {\it Ansatz}
with sufficient enhancement to correctly reproduce the phenomenology of dynamical symmetry
breaking, i.\,e., enhancement of the quark mass function on a domain $p^2\lesssim 1$\,GeV$^2$
and correspondingly a correct value of the chiral condensate. Such an {\it Ansatz} is\cite{Maris:1997tm,Maris:1999nt}
\begin{equation}
\label{calG}
\frac{4\pi\alpha(s)}{s} = \frac{4\pi^2}{\omega^6} \, D\, s\, {\rm e}^{-s/\omega^2}+
\frac{8\pi^2 \gamma_m}{\ln\left[ \tau + \left(1+s/\Lambda_{\rm QCD}^2\right)^2\right]} \, {\cal F}(s)\,,
\end{equation}
with ${\cal F}(s)= [1-\exp(-s/[4 m_t^2])]/s$, $m_t=0.5\,$GeV, $\ln(\tau+1)=2$,
$\gamma_m=12/25$ and $\Lambda_{\rm QCD} = \Lambda^{(4)}_{\overline{MS}} = 0.234\,$GeV.
The free parameters of this model are the range $\omega$ and the strength $D$. One feature
of the model is a specific parameter dependence of ground-state meson masses and properties,
namely that they remain constant over a range of $\omega$, if $\omega\,D=$ const.~is satisfied.\cite{Maris:2002mt}
This results in a one-parameter model, where the free parameter $\omega$ is varied in the
interval $[0.3,0.5]$ GeV. In a calculation, $\omega$ and $D$ as well as the current-quark
masses are fixed to pion mass and decay constant as well as the chiral condensate. That
being done, further results are predictions.

\subsection{Chiral symmetry}

One expression of the chiral properties of QCD is the axial-vector Ward-Takahashi identity
\begin{eqnarray}
P_\mu \Gamma_{5\mu}^j(p;P) = S^{-1}(p_+) i \gamma_5\frac{\tau^j}{2}
+  i \gamma_5\frac{\tau^j}{2} S^{-1}(p_-)- \, 2i\,m(\zeta) \,\Gamma_5^j(p;P)\;,
\label{avwtim}
\end{eqnarray}
which is written here for two quark flavors, each with the same current-quark
mass: $\{\tau^i:i=1,2,3\}$ are flavor Pauli matrices. $\Gamma_{5\mu}^j(k;P)$ and $\Gamma_5^j(k;P)$ are
\begin{figure}[tb]
\centerline{\psfig{file=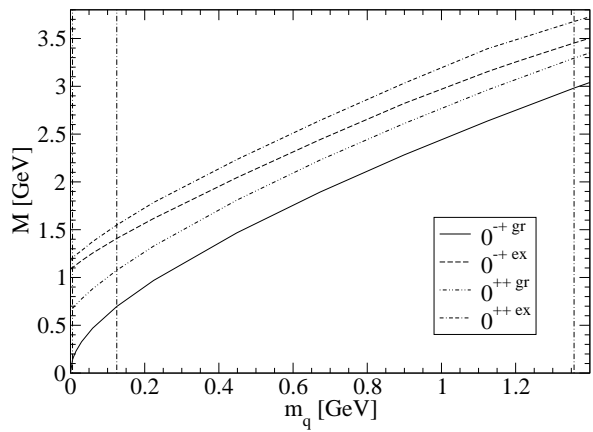,width=5.7cm,angle=270,clip=true}}
\vspace*{8pt}
\caption{Pseudoscalar and scalar meson masses for ground and
excited states as functions of the current quark mass scaled to $1$ GeV by one-loop evolution.
The three vertical dashed-dotted lines (the leftmost
is very close to the $M$-axis) indicate
the values for the $u/d$, $s$, and $c$ quark masses. \label{mesonmasses}}
\end{figure}
the axial-vector and pseudoscalar vertices (for details, see Ref.~\refcite{Maris:1997tm}).
Equation (\ref{avwtim}) is satisfied by relating the kernels of the Bethe-Salpeter and gap equations
(\ref{bse}) and (\ref{gap}), e.\,g., (\ref{ladderK}) and (\ref{rainbowdse}).
A direct consequence of this identity is the relation
\begin{equation}\label{massformula}
f_\mathrm{PS}\,m_\mathrm{PS}^2=2m(\zeta)\rho_\mathrm{PS}(\zeta)\;,
\end{equation}
which relates the pseudoscalar meson mass $m_\mathrm{PS}$ and decay constant $f_\mathrm{PS}$ to
the current quark mass $m(\zeta)$
and the residue of the pseudoscalar vertex at the pion pole $\rho_\mathrm{PS}(\zeta)$ at the
renormalization point $\zeta$. It has been shown\cite{Holl:2004fr} that the implications of Eq.~(\ref{massformula})
in the chiral limit are different for the pion ground and excited states; namely, in the presence of
dynamical chiral symmetry breaking the mass of the ground state vanishes, whereas for the excited states
it is the leptonic decay constant that vanishes instead.

\section{Mesons}

\subsection{Meson ground states}

Ground state mesons and their properties have been studied in rainbow-ladder
truncation using different {\it Ans\"atze} of various sophistication for the effective running
coupling\cite{Maris:1997tm,Maris:1999nt,Munczek:1983dx,Munczek:1991jb,Alkofer:2002bp} over
a range of quark masses including the heavy-quark
domain.\cite{Bhagwat:2004hn,Alkofer:2002bp,Krassnigg:2004if,Maris:2005tt,Ivanov:1997yg,Ivanov:1997iu,Ivanov:1998ms}
The {\it Ansatz} of Ref.~\refcite{Maris:1999nt}, Eq.~(\ref{calG}), has been used successfully
to calculate a large number of pseudoscalar and vector meson properties. We also
used this {\it Ansatz}, since it has the correct ultraviolet behavior and thus yields
reliable results not only for spectroscopy, but also for dynamical observables like
form factors.

\subsection{Radial meson excitations}

It was natural to study radial excitations of pseudoscalar mesons first. An estimate
of the excited pion mass and leptonic decay constant\cite{Krassnigg:2003wy} and the
structure of excited-state Bethe-Salpeter amplitudes, which shows similar characteristics
to a quantum mechanical wave function\cite{Krassnigg:2003dr} were followed by more
detailed studies of pseudoscalar meson radial excitations \cite{Holl:2004fr}
and their electromagnetic properties.\cite{Holl:2005vu} The present work includes
a study of scalar mesons and their first radial excitations.

\subsection{Masses}

Figure \ref{mesonmasses} shows the ground and first radially excited states of pseudoscalar
\begin{figure}[tb]
\centerline{\psfig{file=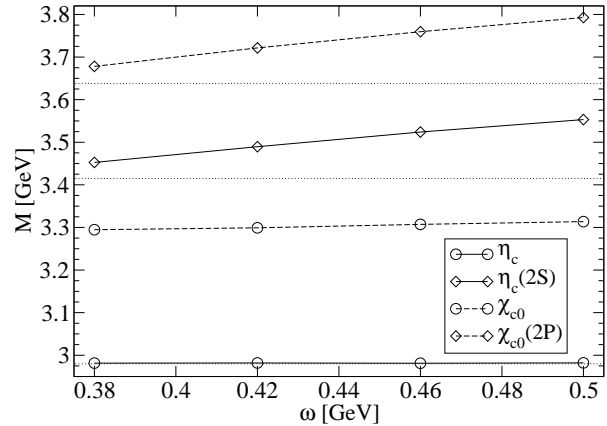,width=5.7cm,angle=270,clip=true}}
\vspace*{8pt}
\caption{Pseudoscalar and scalar $\bar{c}c$ meson masses as
functions of the model parameter $\omega$. The three dotted lines correspond
to the experimental values for the lower three states.
The $\eta_c$ ground-state mass was fitted to the experimental value to
fix the value of the charm-quark mass.\label{charmomegadependence}}
\end{figure}
and scalar mesons as functions of the current-quark mass. Figs.~\ref{mesonmasses} and
\ref{piondecayconstants} are generated from results for $\omega=0.38$ GeV.
In contrast to the ground states, the masses and properties of
radial excitations do depend on the value of $\omega$, even if $\omega\,D=$ const. Since
$r=1/\omega$ corresponds to a range of the infrared part of the interaction, this means
that radial meson excitations provide a means to study the long-range part of the
strong interaction.

To obtain results independent of the choice for $\omega$, one can make use of ratios of
calculated observables, which remain constant over a range of $\omega$, to estimate
properties of states on the basis of experimental values of other states that are known.
An example for such an estimate is that of the mass of the $K(1460)$, the $K$ radial excitation,
where the ratio $M_{K_{ex}}/M_{\pi_{ex}}$ is calculated to be $1.167$; using the experimental
number $M_{\pi_{ex}}=1.3\pm 0.1$ GeV this yields $M_{K_{ex}}=1.52\pm 0.12$ GeV.\cite{Holl:2004un} Meanwhile, we have
performed the analogous calculation for the leptonic decay constants of these states and
found $f_{K_{ex}}/f_{\pi_{ex}}\simeq 10$ in agreement with an estimate via sum rules.\cite{Maltman:2001sv,Maltman:2001gc}

Figure \ref{charmomegadependence} illustrates the same procedure to estimate the mass of the first
scalar radial $\bar{c}c$ excitation $\chi_{c0}(2P)$. The ratio of the calculated masses for
the $\chi_{c0}(2P)$ to the $\eta_c(2S)$ is $1.066$. Via the
experimental value of $3.64$\,GeV (all experimental data are taken from Ref.~\refcite{pdg2006})
for the mass of the $\eta_c(2S)$ we predict the
$\chi_{c0}(2P)$ mass to be $3.88$ GeV. This compares well to quark-model
predictions,\cite{Godfrey:1985xj,Ebert:2002pp,Barnes:2005pb,Lakhina:2006vg}
which lie somewhat below the corresponding estimates from lattice QCD.\cite{Okamoto:2001jb,Chen:2000ej}
We note that the $\chi_{c0}(2P)$ has not yet been observed experimentally.

\subsection{Decay constants}
For the pseudoscalar ground and first radially excited states we plot $f_\mathrm{PS}$ and
\begin{figure}[tb]
\centerline{\psfig{file=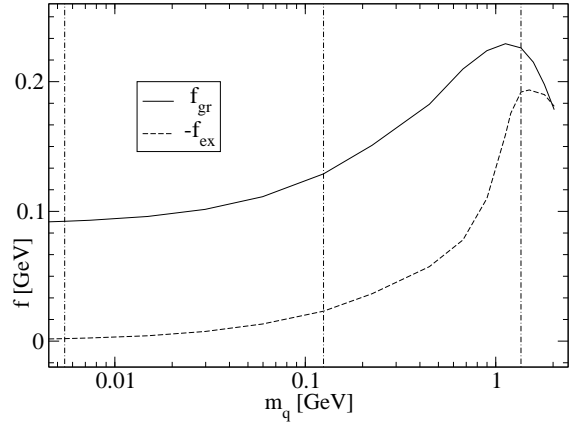,width=5.7cm,angle=270,clip=true}}
\vspace*{8pt}
\caption{$f_\mathrm{PS}$ for the ground- and $-f_\mathrm{PS}$ for the first radially excited-state pseudoscalar
as functions of the current-quark mass scaled to $1$ GeV by one-loop evolution. The three vertical dashed-dotted lines
correspond to the $u/d$, $s$, and $c$ quark masses. Estimated numerical
errors are below $5\%$.\label{piondecayconstants}}
\end{figure}
$-f_\mathrm{PS}$, respectively, as functions of the current-quark mass
in Fig.~\ref{piondecayconstants}. While $f_\mathrm{PS}$ is not necessarily directly
accessible experimentally for $\bar{Q}Q$ systems, where $Q$ is a heavy-quark, it is always a
well-defined axial-vector moment of the meson's Bethe-Salpeter amplitude. Its evolution
with current-quark mass is therefore a useful tool with which to probe QCD and models thereof.
One can see that both curves have a maximum at about the $c$-quark mass;
for higher quark masses, the size of $f_\mathrm{PS}$ decreases for both ground
and excited states. It is remarkable that this ``turning point'' occurs around the same
quark mass as for the heavy-light case.\cite{Maris:2005tt,Ivanov:1998ms}
At the values for the $u/d$-, $s$-, and $c$-quark masses we extract the values for
$f_\mathrm{PS}$ for the ground and $-f_\mathrm{PS}$ for excited states as well as their ratio. The results
are summarized in Table \ref{decayconstanttable}. We note here that for the $s$ quark
our pseudoscalar ground state does not correspond to an actual meson, since it
\begin{table}[tb]
\tbl{Values and ratios of $f_\mathrm{PS}$ for ground- and excited-state pseudoscalar
mesons for $u/d$, $s$, and $c$ quark masses. Current-quark masses are
given at the renormalization point $\zeta$ and scaled to $1$ GeV by one-loop evolution.
All values in GeV.}
{\begin{tabular}{@{}cccccc@{}} \toprule
$\quad$Quark$\quad$ & $m(\zeta=19\;\mathrm{GeV})$ & $\quad m(1\;\mathrm{GeV}) \quad$ &
 $\quad f_\mathrm{gr} \quad$ & $\quad -f_\mathrm{ex} \quad$ &
$\quad -f_\mathrm{ex}/f_\mathrm{gr} \quad$ \\ \colrule
$u/d$ & 0.0037 & 0.00545 & 0.092 & 0.0015 & 0.016 \\
$s$ & 0.0835 & 0.125\hphantom{00} & 0.13\hphantom{0} & 0.023\hphantom{0} & 0.18\hphantom{0} \\
$c$ & 0.905\hphantom{0} & 1.357\hphantom{00} & 0.23\hphantom{0} & 0.19\hphantom{00} & 0.83\hphantom{0}\\ \botrule
\end{tabular} \label{decayconstanttable}}
\end{table}
consists merely of an $\bar{s}s$ component. However, a radial $\bar{s}s$ excitation
can be identified with the $\eta(1475)$.\cite{Holl:2004un} Furthermore,
the ground-state $\bar{s}s$ properties can
also be studied on the lattice, where recent efforts have begun to study ratios of
ground- and excited-state leptonic decay constants.\cite{McNeile:2006qy}

\section{Conclusions and Outlook}

We have extended previous studies of radial meson excitations by studying scalar
excitations and quark masses up to the charm quark. While in the model we used
ground-state meson properties do not depend on variations of the model parameter
$\omega$, the specific parameter dependence of radial excitation properties allows
investigations of the long-range part of the strong interaction between quarks. Without
fixing $\omega$ to a particular value, we used ratios of properties of different excited
states to make estimates for, e.\,g., the leptonic decay constant of the $K(1460)$
and the mass of the $\chi_{c0}(2P)$. This is made an efficacious procedure by the fact
that ratios of excited-state properties remain constant over the domain of $\omega$ under
investigation to a very good level of approximation. For both the ground and excited
equal-mass pseudoscalar states we have calculated $f_\mathrm{PS}$ and observed
a rise to a maximal size around the charm-quark mass and a decrease for higher quark masses.
Further efforts in this direction will include studies of the radial excitations of vector mesons.

\section*{Acknowledgments}

A.~K. is grateful to his colleagues at the Physics Division of Argonne
National Laboratory for their hospitality during a research visit,
where part of this work was completed. We acknowledge useful discussions
with M.\,S.~Bhagwat and A.~H\"oll. This work was supported by:
the Austrian Science Fund FWF, Schr\"odinger-R\"uckkehrstipendium R50-N08;
the Department of Energy, Office of Nuclear
Physics, contract no.~W-31-109-ENG-38; Helmholtz-Gemeinschaft Virtual Theory Institute
VH-VI-041; and benefited from the facilities of ANL's Computing Resource Center.

%\begin{thebibliography}{000} %for 3 digits
%\begin{thebibliography}{00}  %for 2 digits

\end{document}